\documentstyle[aps,times,epsf]{revtex}
\newcommand{\tfrac}[2]{{\textstyle\frac{#1}{#2}}}
\begin{document} 

\title{Derivative Expansion of One-Loop Effective Energy
of Stiff Membranes with Tension}
\draft 
\author{M. E. S. Borelli\thanks{E-mail: borelli@physik.fu-berlin.de},
H. Kleinert\thanks{E-mail: kleinert@physik.fu-berlin.de; website:
http://www.physik.fu-berlin/$\sim$kleinert}, and Adriaan
M. J. Schakel\thanks{E-mail: schakel@physik.fu-berlin.de}} \address{Institut
f\"ur Theoretische Physik \\ Freie Universit\"at Berlin \\ Arnimallee 14,
14195 Berlin.}  \date{\today} \maketitle
\begin{abstract}
With help of a derivative expansion, the one-loop corrections to the
energy functional of a nearly flat, stiff membrane with tension due to
thermal fluctuations are calculated in the Monge parametrization.
Contrary to previous studies, an arbitrary tilt of the surface is
allowed to exhibit the nontrivial relations between the different,
highly nonlinear terms accompanying the ultraviolet divergences.  These
terms are shown to have precisely the same form as those in the original
energy functional, as necessary for renormalizability.  Also infrared
divergences arise.  These, however, are shown to cancel in a nontrivial
way.
\end{abstract}
\vspace{.3cm}
Due to their small surface tension, fluid membranes (see Ref.\
\cite{Jerusalem} for reviews) are subject to strong thermal
undulations.  
The energy of such a membrane is usually modeled by
the local expression \cite{canham,helf1}
\begin{equation} \label{helfmodel}
E_0 = \mu_0 \int {\rm d}S + \frac{1}{2} \int {\rm d}S (\kappa_0H^2 +
\bar{\kappa}_0 K ), 
\end{equation} 
where ${\rm d}S$ are the surface elements, while $H$ and $K$ denote (twice) the mean
and the Gaussian curvature of the membrane surface, respectively.  In terms of
the principal radii $R_1$ and $R_2$ of curvature these are
$H = 1/R_1 + 1/R_2$ and $K = 1/R_1 R_2$.  The parameter
$\mu_0$ in (\ref{helfmodel}) is the surface tension, $\kappa_0$ the
bending rigidity, and $\bar{\kappa}_0$ its Gaussian counterpart.  The
geometric quantities appearing in the integral (\ref{helfmodel}) are
invariant under translations and rotations in space.  They are also
independent of the parametrization of the surface.  We ignore a possible
spontaneous curvature term linear in $H$.  The energy of a physical membrane
contains all higher powers in the principal radii of curvature, but these are
irrelevant at large length scales.  (In the language of renormalization
group analysis, the first term is relevant, the second and third are marginal.)

The statistical behavior of fluctuating membranes was first studied by
Helfrich \cite{helf1} using only curvature terms.  We allow for an additional
surface term in (\ref{helfmodel}) because flucutations arising from the
curvature terms not only renormalize the bending rigidities, but also the
tension.  The thermal fluctuations soften the bending rigidity at large
length scales, reducing it from the bare value $\kappa_0$ as follows
\begin{equation}  \label{difkappas}
\kappa_{\rm{eff}} = \kappa_0 - \frac{\alpha}{4 \pi \beta} \ln (\Lambda L),
\end{equation} 
where $\beta$ is the inverse temperature, $\Lambda$ is an ultraviolet
momentum cutoff of the order of the inverse microscopic length scale $a$
given by the length of the molecules, whereas $L$ is an infrared cutoff
determined by the finite size of the membrane.  Various authors derived
different values for $\alpha$, first $\alpha = 1$
\cite{helf3,helf2,forster2} was obtained, later $\alpha = 3$
\cite{peliti,forster,klein1,fluid}.  The second result has also been
found in computer simulations \cite{gompper}.  For either value of
$\alpha$, the rigidity disappears at length scales larger than the
persistence length \cite{DGT}
\begin{equation} 
\xi \sim a \exp  \left(\frac{4 \pi}{\alpha} \beta \kappa_0 \right),
\end{equation} 
beyond which the normal vectors of the surface become uncorrelated---the
surface looks crumpled.  More recent calculations \cite{helf4} suggest
the value $\alpha = -1$, implying a stiffening instead of a softening of
the bending rigidity.  This new result was argued to arise from the use
of another integration measure which respects the incompressible-fluid
nature of the membrane from the outset.  This is in contrast to previous
studies of in-plane fluid \cite{fluid} and elastic effects \cite{elast}
which did not show any change in the value $\alpha=3$ (they only enter
at the two-loop level \cite{davidJ}).

The renormalization of the Gaussian rigidity $\bar \kappa_0$ was first
calculated in \cite{klein1} to have the same form as in (\ref{difkappas}),
but with $\alpha \rightarrow \bar{\alpha} = -10/3$.  This value {\em is}
changed by in-plane fluid and elastic effects \cite{fluid,elast}.

The renormalization group flow of $\kappa_{\rm eff}$ extracted from the
one-loop result (\ref{difkappas}) has no nontrivial fixed point.  If
this conclusion persists to all orders in perturbation theory, it would
imply the absence of a smooth phase with long-ranged correlations.  The
smooth appearance of lipid vesicles in the laboratory can then only be
explained by their very large persistence length.  An alternative
explanation has recently been proposed in \cite{klein5}, where it was
argued that the neglected higher order terms in the energy
(\ref{helfmodel}) may give rise to a nonperturbative mechanism, by which
the crumpled phase can go over into a smooth phase via a sequence of two
Kosterlitz-Thouless phase transitions or a single first-order one. 

The renormalization of the surface tension has also been investigated by
several authors.  The results can be summarized by the formula
\begin{equation} \label{difmus} 
\mu_{\rm{eff}} = \mu_0 + \frac{\alpha'}{4 \pi \beta}
\frac{\mu_0}{\kappa_0} \ln (\Lambda L ),
\end{equation} 
with the value $\alpha' = 1$ found in \cite{forster,david} and $\alpha' = 3$
in \cite{peliti,klein1}.  In Ref.\ \cite{cai}, an attempt was made to
reconcile the differences.  An almost planar surface without overhangs was
considered in the Monge parametrization.  The points on the surface are then
specified by a vector field ${\bf r}(x) = (x_1,x_2,\phi(x))$, where
$\phi(x)$ denotes the vertical displacement of the surface with respect to a
base plane with Cartesian coordinates $x=(x_1,x_2)$.  For a surface with
fixed topology, the Gaussian curvature energy is a constant and can be
ignored.  The remainder of Eq.\ (\ref{helfmodel}) was expanded to fourth
order in the displacement field.  The relative weights of the resulting
terms are fixed by their covariant origin.  The authors encountered
considerable problems in showing that this remains true after including the
thermal fluctuations.  They studied the renormalization of the surface
tension by determining the coefficient $\tau$ of the first (constant) term
in the expansion of the surface energy,
\begin{eqnarray} \label{tau}
\tau &=& \mu_0 + \frac{1}{2 \beta} \int \frac{{\rm d}^2 k}{(2 \pi)^2}
\ln(\mu_0 k^2 + \kappa_0 k^4) \nonumber \\
     &=& \mu_0 + \frac{1}{4 \pi \beta} \frac{\mu_0}{\kappa_0}
\ln(\Lambda L) + c_1 \Lambda^2 + c_2 \Lambda^2
\ln \Lambda + c_3,
\end{eqnarray}
with $c_i$ constants, and comparing it with the coefficient $\mu_{\rm
eff}$ of the second term [proportional to $(\partial \phi)^2$],
\begin{eqnarray} \label{mu}
\mu_{\rm eff} &=& \mu_0 - \frac{1}{2 \beta} \int \frac{{\rm d}^2 k}{(2
\pi)^2} \left[ 3 - \frac{\mu_0 k^2}{\mu_0 k^2 + \kappa_0 k^4} \right]
\nonumber \\ &=& \mu_0 + \frac{1}{4 \pi \beta} \frac{\mu_0}{\kappa_0}
\ln(\Lambda L) + c_4 \Lambda^2 + c_5.
\end{eqnarray}
The above expressions differ from each other by positive powers of the
cutoff $\Lambda$.  Since the covariance of the theory implies $\tau =
\mu_{\rm eff}$, two factors were added to the energy, in order to correct
Eq.\ (\ref{mu}).  The first one corresponds to the Faddeev-Popov determinant
associated with fixing the gauge.  In the Monge gauge, its contribution is
proportional to the cutoff $\Lambda$.  The second (more ad hoc) factor
introduces a nonlinear correction to the integration measure in the
partition function.  It accounts for the difference between an infinitesimal
surface element on the membrane and its projection on the reference plane.
This second correction factor, too, contributes only with positive powers of
the cutoff.  Added to the first one, it leads to the equality $\mu_{\rm eff}
= \tau$, which is the main result of Ref.\ \cite{cai}.

Motivated by these problems, and by the renewed interest in the subject,
we study the role of thermal fluctuations in a more general approach.
Employing a derivative expansion \cite{caroline}, we calculate the full
effective energy functional produced by Gaussian fluctuations for an
arbitrary background configuration, maintaining the full nonlinear
structure of the energy at all intermediate steps.

The mean curvature in the Monge parametrization reads
\begin{equation} 
H=  \partial \cdot N =  \partial_\mu N_\mu,
\end{equation}  
where the summation is over the first two components only $(\mu =1,2)$;
${\bf N}$ is the unit normal to the surface
\begin{equation} 
{\bf N} = \frac{1}{\sqrt{1 + (\partial \phi)^2}}\left(-\partial_1 \phi,
-\partial_2 \phi, 1 \right),
\end{equation}
and the surface elements are
\begin{equation} 
{\rm d}S =  {\rm d}^2x \sqrt{1 + (\partial \phi)^2},
\end{equation} 
so that the first two terms of the energy (\ref{helfmodel}) read
explicitly
\begin{equation} \label{themodel}
E_0[\phi] = \int {\rm d}^2x \sqrt{1+(\partial \phi)^2} \Biggl\{
\mu_0 + \frac{\kappa_0}{2} \biggl[ \frac{(\partial^2
\phi)^2}{1+(\partial \phi)^2} - 2 \frac{\partial_\mu \phi \partial_\nu
\phi \partial_\mu \partial_\nu \phi \partial^2 \phi}{[1+(\partial
\phi)^2]^2} + \frac{(\partial_\mu \phi \partial_\nu \phi \partial_\mu
\partial_\nu \phi)^2}{[1+(\partial \phi)^2]^3} \biggr] \Biggr\} .
\end{equation} 
Physically, $\mu_0$ corresponds to the chemical
potential specifying the exchange of molecules between the
(incompressible) membrane and its aqueous environment.

The main purpose of this note is to show that the ultraviolet divergent parts of the
one-loop corrections
induced by thermal fluctuations are of precisely the same form as in
(\ref{themodel}), and in particular, that the three terms in the curvature energy
renormalize in the same way, resulting in an overall renormalization of
$\kappa_0$ alone.

To apply the derivative expansion we write the partition function as a
functional integral over the displacement field
\begin{equation}  \label{partfunc}
Z = \int {\rm D} \phi \exp \left( - \beta E_0 \right),
\end{equation} 
with each field configuration weighted with a Boltzmann factor.  Fixing a
gauge is generally accompanied by a Faddeev-Popov determinant appearing in
the measure of the functional integral.  Following Ref.\ \cite{leibler} we
adopt dimensional regularization to handle momentum integrals which diverge
in the ultraviolet.  This is common practice in the technically closely
related nonlinear sigma model.  The great advantage of dimensional
regularization over regularization with a momentum cutoff $\Lambda \sim
1/a$, is that terms diverging with a strictly positive power of $\Lambda$
are suppressed.  As a result, both the Faddeev-Popov determinant
corresponding to the Monge gauge and the second correction factor introduced
in Ref.\ \cite{cai}, which contain only positive powers of the cutoff, are
unity in dimensional regularization, and the difficulties addressed in that
reference are avoided.  Only logarithmic divergences show up as poles in
$\epsilon$, where $\epsilon = 2 - D$, $D$ being the dimension of the
membrane.  The connection between the two types of regularization is
\begin{equation}  \label{rel}
\frac{1}{\epsilon} \rightarrow \ln (\Lambda L);
\end{equation} 
with the linear size $L$ of the membrane representing the relevant
long-distance scale.  The rationale for using dimensional
regularization is that contributions to the effective energy with 
strictly positive powers of the ultraviolet cutoff are connected to
 $\delta^{(2)}(x=0)$. 
These highly local terms are
uninteresting at large lenght scales \cite{CaoSc,Donoghue}.

In the one-loop approximation, the exponent in (\ref{partfunc}) may be
expanded up to second order around a background configuration
$\Phi(x)$ extremizing $E_0$.  A nontrivial background requires the
presence of an extra source term.  For brevity, this term will not be
written down explicitly when setting $\delta E_0 / \delta \Phi = 0$.
The resulting integral is Gaussian and yields an
effective energy
\begin{equation} \label{heff1}
E_{\rm eff}[\Phi] = E_0 [\Phi] + E_1 [\Phi] = E_0 [\Phi] + \frac{1}{2 \beta}
\mbox{Tr} \ln \left[\left. \frac{\delta^2 (\beta E_0)}{\delta \phi(x) \delta
\phi(y)} \right|_{\Phi} \right],
\end{equation} 
where the expression in square brackets corresponds to the matrix of
second functional derivatives of $E_0$ and the trace Tr stands for the
trace of this matrix, i.e., the integral $\int {\rm d}^2 x$ over
space, as well as the integral $\int {\rm d}^2 k/(2 \pi)^2$ over
momentum \cite{caroline}.

The one-loop correction $E_1 [\Phi]$ to the energy will now be calculated in
a derivative expansion for a nearly flat, but arbitrarily tilted background
configuration. The expansion has the general form
\begin{equation} \label{heff2}
E_1 [\Phi] = \int {\rm d}^2x \left[ {\cal V}(V_\lambda) + {\cal Z}^1(V_\lambda)
(\partial_\mu V_\mu)^2 + {\cal Z}^2_{\mu \nu}(V_\lambda) \partial_\mu V_\nu
\partial_\sigma V_\sigma + {\cal Z}^3_{\mu \nu \sigma \rho}(V_\lambda)
\partial_\mu V_\nu \partial_\sigma V_\rho + \cdots \right],
\end{equation} 
where we introduced the abbreviation $V_\mu = \partial_\mu \Phi$, while
${\cal V}$, ${\cal Z}^1$, ${\cal Z}^2_{\mu \nu}$, and ${\cal Z}^3_{\mu \nu
\sigma \rho}$ are functions of $V_\mu$ to be determined.  Following Ref.\
\cite{caroline}, we set $V_\mu(x) = {\bar V}_\mu + v_\mu(x)$, where ${\bar
V}_\mu$ denotes the constant part of $V_\mu(x)$, and expand Eq.\
(\ref{heff2}) in powers of $v_\mu(x)$ and its derivatives, to obtain
\begin{eqnarray} \label{exp}
E_1 [\bar{V}_\lambda + v_\lambda] = \int {\rm d}^2x \biggl[ && {\cal
V}(\bar{V}_\lambda) + \frac{\partial {\cal V}
(\bar{V}_\lambda)}{\partial \bar{V}_\mu } v_\mu + \frac{1}{2}
\frac{\partial^2 {\cal V}(\bar{V}_\lambda)}{\partial \bar{V}_\mu
\partial \bar{V}_\nu } v_\mu v_\nu + {\cal Z}^1(\bar{V}_\lambda)
(\partial_\mu v_\mu)^2 \nonumber \\ && + {\cal Z}^2_{\mu
\nu}(\bar{V}_\lambda) \partial_\mu v_\nu \partial_\sigma v_\sigma +
{\cal Z}^3_{\mu \nu \sigma \rho}(\bar{V}_\lambda) \partial_\mu v_\nu
\partial_\sigma v_\rho + \cdots \biggr],
\end{eqnarray} 
with space-independent ${\cal V}(\bar{V}_\lambda)$ and ${\cal Z}(\bar{V}_\lambda)$'s.
These functions will now be extracted from the expansion of the Tr ln in 
(\ref{heff1}) up to quadratic terms in $v_\mu$ and $\partial_{\mu} v_\nu$.

The functional derivatives in (\ref{heff1}) are calculated using the
Euler-Lagrange formula
\begin{equation} 
\frac{\delta F[\phi]}{\delta \phi (x)} = \frac{\partial f}{ \partial
\phi} - \partial_\mu \frac{\partial f}{\partial (\partial_\mu \phi)} +
\partial_\mu \partial_\nu \frac{\partial f}{\partial (\partial_\mu
\partial_\nu \phi)} + \cdots \label{funcderiv},
\end{equation} 
with $F[\phi] = \int {\rm d}^2 x f(\phi, \partial_\mu \phi,
\partial_\mu \partial_\nu \phi, \ldots )$.  To keep track of the many
terms appearing in the resulting expression we have used the algebraic
computer program {\tt FORM} \cite{form}.

We consider first the renormalization of the surface tension.
Since the energy density $\sqrt{1 + V^2}$ does not
contain derivatives of $V_\mu$, we may set $v_\mu(x)$ to zero and
consider $E_1[\bar{V}_\lambda]$ only,
\begin{equation} \label{gamma0}
\beta E_1[\bar{V}_\lambda] = - \tfrac{1}{4} \mbox{Tr} \ln
(1+\bar{V}^2) + \tfrac{1}{2} \mbox{Tr} \ln [G^{-1}(p)] .
\end{equation} 
Here, $G^{-1}(p)$ denotes the inverse propagator:
\begin{equation}  \label{prop}
G^{-1}(p) = (\mu_0 + \kappa_0 p^2) p^2 - (\mu_0 + 2 \kappa_0 p^2)(\bar{U} \cdot p)^2
+\kappa_0 (\bar{U} \cdot p)^4
\end{equation} 
where $\bar{U}_{\mu}$ is the constant vector
\begin{equation} 
\bar{U}_{\mu} = \frac{\bar{V}_\mu}{\sqrt{1+\bar{V}^2}}.
\end{equation} 
In dimensional regularization, the first term at the right-hand side of
(\ref{gamma0}) is zero.  To evaluate the remaining ${\rm Tr} \ln$, we
apply a standard trick and first differentiate (\ref{gamma0}) with
respect to $\mu_0$ to obtain
\begin{equation} \label{dgamma}
\frac{\partial E_1[\bar{V}_\lambda]}{\partial \mu_0} =
\frac{1}{2 \beta} \mbox{Tr} \left[ \frac{p^2 - (\bar{U} \cdot p)^2 }{(\mu_0 +
\kappa_0 p^2) p^2 - (\mu_0 + 2 \kappa_0 p^2)(\bar{U} \cdot p)^2 +\kappa_0
(\bar{U} \cdot p)^4} \right] ,
\end{equation} 
where $p_{\mu} = - i \partial_\mu$.  Because the integrand contains no
space-dependence, the spatial part of the trace in (\ref{dgamma}) yields
an area factor $A = \int {\rm d}^Dx$, and we are left with the momentum
integral
\begin{equation} 
\frac{\partial E_1[\bar{V}_\lambda]}{\partial \mu_0} =
\frac{A}{2 \beta} \int \frac{{\rm d}^D k}{(2 \pi)^D} \left[ \frac{k^2
- (\bar{U} \cdot k)^2 }{(\mu_0 + \kappa_0 k^2) k^2 - (\mu_0 + 2 \kappa_0
k^2)(\bar{U}\cdot k)^2 +\kappa_0 (\bar{U} \cdot k)^4} \right] .
\end{equation} 

Being interested only in the ultraviolet divergent terms, we obtain, in
dimensional regularization
\begin{equation} \label{dgamma2}
\frac{\partial E_1[\bar{V}_\lambda]}{\partial \mu_0} =
\frac{1}{4 \pi \kappa_0 \beta} \frac{1}{\epsilon} \int {\rm d}^2x 
\sqrt{1+\bar{V}^2} .
\end{equation} 
After integrating again with respect to $\mu_0$ and comparing the result
with (\ref{heff2}) we find (up to an irrelevant additive constant)
\begin{equation}  \label{z0} 
{\cal V}(V_\lambda) = \frac{\mu_0}{4 \pi \beta \kappa_0}\frac{1}{\epsilon}
\sqrt{1+ V^2} = \frac{\mu_0}{4 \pi \beta \kappa_0}\frac{1}{\epsilon}
\sqrt{1+ (\partial \Phi)^2},
\end{equation}
where we replaced $\bar{V}_\lambda$ with the full background field
$V_\lambda(x)$, to obtain the first term in (\ref{heff2}).  Note that
this one-loop correction is precisely of the same form as the surface
term contained in the original energy expression (\ref{helfmodel}).
This term can consequently be combined with the original one by
introducing the renormalized tension
\begin{equation}  \label{mueff}
\mu_{\rm eff} = \mu_0 + \frac{1}{4 \pi \beta} \frac{\mu_0}{\kappa_0}
\frac{1}{\epsilon}.
\end{equation} 
This result, corresponding to $\alpha' = 1$ in (\ref{difmus}), is in
agreement with Refs.\ \cite{forster,david}, but disagrees with Refs.\
\cite{peliti,klein1} where the value $\alpha' = 3$ was obtained.  To
understand the differences, we note that in these last two references, the
energy (\ref{helfmodel}) with $\mu_0=0$ was used instead.  That is, the
renormalization of the surface tension calculated by these authors was
generated solely by the curvature terms.  However, the surface term also
contributes.  In fact, it generates a contribution with $\alpha'=-2$, which,
together with the contribution obtained from the curvature terms, results in
the value $\alpha' = 1$.  This is also the value obtained in Ref.\
\cite{cai}, as can be seen by disregarding the terms proportional to positive
powers of $\Lambda$ in Eq.\ (\ref{tau}) and using the relation (\ref{rel}).
There, the covariance of the first two terms in the expansion of the surface
energy is assured by introducing correction factors proportional to positive
powers of the cutoff.  Our result, based on dimensional regularization where
terms with positive powers of the cutoff are suppressed, proves the
covariance of {\it all} terms in the expansion of the surface energy since
the full expression has been maintained.

We continue to investigate the renormalization of the bending rigidity.
Since the three terms involved contain derivatives of the background
field $V_\mu$, we now have to employ the derivative expansion.  As a
first step, we Taylor expand the logarithm in (\ref{heff1}) as:
\begin{equation}  \label{traceexpand} 
\beta E_1[\bar{V}_\lambda + v_\lambda(x)] - \beta
E_1[\bar{V}_\lambda] = \tfrac{1}{2} \mbox{Tr} \ln[1 + G(p)
\Lambda(x,p)] = \tfrac{1}{2} \mbox{Tr}[G(p) \Lambda(p,x)]
- \tfrac{1}{4} \mbox{Tr}[G(p) \Lambda(x,p) G(p) \Lambda(x,p)] + \cdots ,
\end{equation}  
where $G(p)$ is the propagator defined in (\ref{prop}) and
$\Lambda(x,p)$ contains the $x$-dependent terms obtained from
functionally differentiating $E_0$ twice, setting $\partial_\mu
\Phi(x) = V_{\mu}(x) = \bar{V}_\mu + v_\mu(x)$ and expanding up to
second order in $v_\mu$ and $\partial_\mu v_\nu$.

The first term in (\ref{traceexpand}) can be calculated in a similar
fashion as ${\cal V}(V_\lambda)$. In the second term, all momentum
operators have to be moved to the left \cite{caroline}, by repeatedly
applying the identity
\begin{equation} \label{commu}
f(x) p_\mu g(x) = (p_\mu + i \partial_\mu) f(x) g(x),
\end{equation} 
where $f(x)$ and $g(x)$ are arbitrary functions and the derivative
$\partial_\mu$ acts {\it only} on the next object to the right, while
the derivative $p_\mu$ acts on {\it everything} to the right.

The typical momentum integrals showing up at the one-loop order are of the
form
\begin{equation} \label{I}
I_{m,n} = \int \frac{{\rm d}^D k}{(2 \pi)^D} k^m
G^n(k) \sim \int {\rm d}k \left\{ \begin{array}{l} k^{m + D -1
- 2 n} \;\;\;\; {\rm infrared} \\ k^{m + D -1
- 4 n} \;\;\;\; {\rm ultraviolet} \end{array} \right. ,
\end{equation} 
with $m, n >0$.  They diverge in the infrared when $m + D -1 - 2 n \leq
-1$, and in the ultraviolet when $m + D -1 - 4 n \geq -1$.  For $D=2$
these conditions become $m - 2 n \leq -2$, $m - 4 n \geq -2$,
respectively, and the two types of divergences are seen to be separated
by a wedge of finite integrals in the $(m,n)$-plane starting at
$(-2,0)$.

After a tedious and lengthy calculation, involving of the order of
$10^4$ terms, done with help of a program written in {\tt FORM} \cite{form}, 
we obtained the divergent terms to second order in derivatives of the
field $v_\mu$:
\begin{eqnarray} \label{results}
\beta E_1[\bar{V}_\lambda + v_\lambda] - \beta
E_1[\bar{V}_\lambda] =  \int {\rm d}^2x \Biggl\{ &&
\frac{\mu_0}{4 \pi \kappa_0} \frac{1}{\epsilon}
\left[\frac{\bar{V}_\mu}{(1 + \bar{V}^2)^{1/2}} v_\mu +
\frac{1}{2} \left( \frac{\delta_{\mu\nu}}{(1 + \bar{V}^2)^{1/2}} +
\frac{\bar{V}_\mu \bar{V}_\nu}{(1 + \bar{V}^2)^{3/2}} \right)
v_\mu v_\nu \right] \nonumber \\ && - \frac{3}{8 \pi}
\frac{1}{\epsilon} \left[\frac{1}{(1 + \bar{V}^2)^{1/2}}(\partial_\mu
v_\mu)^2 - 2 \frac{\bar{V}_\mu \bar{V}_\nu}{(1 + \bar{V}^2)^{3/2}}
\partial_\mu v_\nu \partial_\sigma v_\sigma +
\frac{\bar{V}_\mu \bar{V}_\nu \bar{V}_\sigma \bar{V}_\rho}{(1 +
\bar{V}^2)^{5/2}} \partial_\mu v_\nu \partial_\sigma v_\rho
\right] \nonumber \\ && -  \frac{1}{4 \pi} \frac{1}{\epsilon_{\rm
ir}} \left[\frac{\bar{V}_\mu \bar{V}_\nu}{(1 + \bar{V}^2)^{3/2}}
\partial_\mu v_\nu \partial_\sigma v_\sigma -
\frac{\bar{V}_\mu \bar{V}_\nu \bar{V}_\rho \bar{V}_\sigma}{(1 +
\bar{V}^2)^{5/2}}\partial_\mu v_\nu \partial_\sigma v_\rho
\right] \Biggr\} .
\end{eqnarray} 
In deriving this expression we also encountered infrared divergences.  These
are regularized in the same scheme as used to regularize the ultraviolet
divergences.  To distinguish the two we gave epsilon an index ir in case of
an infrared divergence.  We leave the discussion of the infrared divergences
to the next paragraph, and first analyze the ultraviolet ones.  Comparing
(\ref{results}) to (\ref{exp}) with ${\cal V}(\bar{V}_\lambda)$ given by
(\ref{z0}), we see that the terms proportional to $\mu_0$ precisely
correspond to the first two terms at the right-hand side of (\ref{exp}), as
it should be.  Moreover, we conclude that the ${\cal Z}$-functions in
(\ref{exp}) are given by
\begin{eqnarray} \label{z3}
{\cal Z}^1({\bar V}_\lambda) = -\frac{3}{8 \pi \beta}
\frac{1}{\epsilon} \frac{1}{(1 + {\bar V}^2)^{1/2}}, \, {\cal
Z}^2_{\mu \nu}({\bar V}_\lambda) = \frac{3}{4 \pi \beta}
\frac{1}{\epsilon} \frac{{\bar V}_\mu {\bar V}_\nu}{(1 + {\bar
V}^2)^{3/2}}, \, {\cal Z}^3_{\mu \nu \sigma \rho}({\bar
V}_\lambda) = - \frac{3}{8 \pi \beta} \frac{1}{\epsilon} \frac{{\bar
V}_\mu {\bar V}_\nu {\bar V}_\sigma {\bar V}_\rho}{(1 + {\bar
V}^2)^{5/2}}.
\end{eqnarray} 
By replacing the constant $\bar{V}_\lambda$ with the
full background field $V_\mu(x) = \partial_\mu \Phi(x)$, 
we obtain for the divergent parts of the expansion 
(\ref{heff2}) the explicit form
\begin{equation} 
\beta E_1 [\Phi] = \frac{1}{4 \pi \epsilon \kappa_0} \int {\rm d}^2x
\sqrt{1+(\partial \Phi)^2} \Biggl\{ \mu_0  - \frac{3 \kappa_0}{2}
\biggl[ \frac{(\partial^2 \Phi)^2}{1+(\partial \Phi)^2} - 2
\frac{\partial_\mu \Phi \partial_\nu \Phi \partial_\mu \partial_\nu \Phi
\partial^2 \Phi}{[1+(\partial \Phi)^2]^2} + \frac{(\partial_\mu \Phi
\partial_\nu \Phi \partial_\mu \partial_\nu \Phi)^2}{[1+(\partial
\Phi)^2]^3} \biggr] \Biggr\} .
\end{equation} 
We see that the thermally generated terms at the one-loop level are
precisely of the same form as those present in the original energy
expression (\ref{themodel}).  In addition, the relative weights of the
curvature terms produced by the fluctuations are the same as those found
there.  They can therefore be combined with the original terms by
introducing the renormalized rigidity
\begin{equation} \label{kappaeff}  
\kappa_{\rm eff} = \kappa_0 - \frac{3}{4 \pi \beta}\frac{1}{\epsilon},
\end{equation}  
whose value is in agreement with
\cite{peliti,forster,klein1,fluid,gompper}.

As seen in (\ref{results}), the one-loop corrections seem to have
introduced infrared divergences in the theory.  A closer inspection
reveals that the infrared-divergent contributions all stem from the
surface energy term in (\ref{themodel}), so that it suffices to analyze
the one-loop corrections to the truncated energy
\begin{equation}  \label{hammu}
E'_0 = \mu_0 \int {\rm d}^2x \sqrt{1+(\partial \phi)^2}.
\end{equation} 
Infrared divergences in this model have previously been studied in
\cite{david2}, where they were shown to disappear for an infinitely
small dimension $D$ of the membrane to all orders in $D$.  In our
calculation the problem arises for $D = 2 - \epsilon$.  When calculating
the effective action, we expand (\ref{hammu}) around the background
field $\Phi$ extremizing $E'_0$, i.e.,
\begin{equation} \label{fieldeq1}
\left. \frac{\delta E'_0}{\delta \phi} \right|_{\Phi} = 0,
\end{equation} 
which reads explicitly
\begin{equation} 
\frac{\partial^2 \Phi}{[1 + (\partial \Phi)^2]^{1/2}} -
\frac{\partial_\rho \Phi \partial_\sigma \Phi \partial_\sigma \partial_\rho
\Phi}{[1 + (\partial \Phi)^2]^{3/2}} = 0 \label{fieldeq2}. 
\end{equation}
The presence of the implicitly assumed sources turns this equation in a
nontrivial one.  Rewriting $\partial_\mu \Phi(x) = \bar{V}_{\mu} +
v_{\mu}(x)$, expanding to linear order in $v_\mu$, and substituting the
resulting expression in (\ref{results}), we see the infrared divergences
to vanish for a two-dimensional membrane.

In conclusion, we have demonstrated that all logarithmically divergent
one-loop corrections induced by thermal fluctuations are precisely of
the same form as in the original
energy (\ref{themodel}), so that they can be removed by a
renormalization of the surface tension and bending rigidity.

\acknowledgments 
We thank V. Schulte-Frohlinde and S. Shabanov for useful discussions, and
W. Helfrich and H. A. Pinnow for sharing their results prior to
publication.

\end{document}